# GRAPHIC – Guidelines for Reviewing Algorithmic Practices in Human-centred Design and Interaction for Creativity

Joana Rovira Martins*

University of Coimbra, CISUC/LASI – Centre for Informatics and Systems of the University of Coimbra, Department of Informatics Engineering; University of Coimbra, Interdisciplinary Research Institute, Computational Media Design, jmmartins@dei.uc.pt

Pedro Martins

University of Coimbra, CISUC/LASI – Centre for Informatics and Systems of the University of Coimbra, Department of Informatics Engineering, pjmm@dei.uc.pt

Ana Boavida

University of Coimbra, CISUC/LASI – Centre for Informatics and Systems of the University of Coimbra, Department of Informatics Engineering, aboavida@dei.uc.pt

Artificial Intelligence (AI) has been increasingly applied to creative domains, leading to the development of systems that collaborate with humans in design processes. In Graphic Design, integrating computational systems into co-creative workflows presents specific challenges, as it requires balancing scientific rigour with the subjective and visual nature of design practice. Following the PRISMA methodology, we identified 872 articles, resulting in a final corpus of 71 publications describing 68 unique systems. Based on this review, we introduce GRAPHIC (Guidelines for Reviewing Algorithmic Practices in Human-centred Design and Interaction for Creativity), a framework for analysing computational systems applied to Graphic Design. Its goal is to understand how current systems support human-AI collaboration in the Graphic Design discipline. The framework comprises main dimensions, which our analysis revealed to be essential across diverse system types: (1) Collaborative Panorama, (2) Processes and Modalities, and (3) Graphic Design Principles. Its application revealed research gaps, including the need to balance initiative and control between agents, improve communication through explainable interaction models, and promote systems that support transformational creativity grounded in core design principles.

**CCS CONCEPTS** • Applied computing → Media arts • Applied computing → Computer-aided design • General and reference → Design • Human-centered computing → Collaborative interaction

**Additional Keywords and Phrases:** Human-Computer Co-creation, Graphic Design, Creative Systems, Design Principles

---

* Corresponding author.

# 1 INTRODUCTION

Computational approaches can be explored to amplify human creative potential, as seen in human-computer co-creative systems, defined by their ability to collaborate with humans to generate new ideas [35, 38]. Despite advances in the field, these systems still present challenges due to the unpredictability of human creativity and the diverse strategies that can be employed during the creative process [69]. Human behaviour can influence the creative process to the extent that the need to control it can vary both from use to use and within the various phases of the same process [69]. Therefore, beyond technical skills, the development of such tools should consider the subjective nature of creative cognition and the interaction models that effectively support co-creation [44, 64].

In design, this challenge is complex since it requires balancing scientific rigour with design nuances, two notions often considered epistemologically opposed [44]. This issue is particularly pronounced in Graphic Design, which relies on visual language to communicate effectively. One way to address this is by considering a basic visual vocabulary of communication, i.e., the fundamental principles of Graphic Design [45]. These (such as balance, emphasis, or contrast) serve as a foundation for developing visually effective and comprehensible solutions [68]. As Leborg (2013) argues, reflecting on our creations alters the creative process, since we think differently when we have a language to describe things. Accordingly, co-creative systems in this context should facilitate experimentation and support clear reflection on design principles used.

We conducted a systematic literature review on computational systems in Graphic Design. Following the Preferred Reporting Items for Systematic reviews and Meta-Analyses (PRISMA) methodology [60], we analysed 872 articles, of which 25 were included. We also added 46 articles to our final corpus based on citations, totalling 71 articles analysed, describing 68 unique systems—three systems were discussed in six separate publications. This overview aims to enhance researchers' understanding of the current state of co-creative systems in Graphic Design. We propose the Guidelines for Reviewing Algorithmic Practices in Human-centred Design and Interaction for Creativity (GRAPHIC) framework. This is based on three dimensions: (1) Collaborative panorama, which analyses human-computer partnership; (2) Explored modalities and processes, encompassing both technical and creative aspects; and (3) Graphic Design principles, which ground the analysis in the discipline's visual language. These dimensions have allowed us to identify potential research opportunities that go beyond this framework and can be adapted by adjacent design areas.

# 2 BACKGROUND

Artificial Intelligence (AI) skills are not the only decisive factor in the success of a human-AI collaboration [54]. Various researchers have proposed guidelines and frameworks to understand and enhance interaction models [1, 2, 51, 65, 71]. Regarding interaction, Shi et al. [77] developed a taxonomy of Human-GenAI interaction divided into five dimensions: Purposes of Using GenAI, Feedback from Models to Humans, Control from Humans to Models, Levels of Engagement, and Application Domains. More recently, Rezwana and Ford [70] developed FAICO, a framework that identifies key aspects of effective human-AI communication, including modalities, Response Mode, Timing, Types, Explanation Details, and Tone.

Within the scope of shared control between humans and AI, Cimolino and Graham [14] proposed an analytical framework that confronts six domains across four axes: Role of AI, Supervision, Influence and Mediation. Moruzzi and Margarido [59] introduced the UCCC framework, defining nine dimensions organised into three categories: Interaction Guidance & Response, Interaction Configuration, and Interaction Dynamics.

These methods are characterised by their broad orientation. By focusing on specific areas of design, different challenges are identified, such as the ability to adapt to diverse user needs [69]. Windl et al. [90] identified four strategies that describe how designers include AI models into their design process: a priori, post-hoc, model-centric, and competence-centric. Another example that helps characterise the designer-AI partnership is the approach proposed by Shi et al. [78], which defines five dimensions: Scope, Access, Agency, Flexibility, and Visibility. Hwang [25] also analysed various AI-based design support tools and identified four categories: Editors, Transformers, Blenders, and Generators. Despite these contributions, significant gaps remain, particularly in fields like Graphic Design. To address this, we propose a framework based on a systematic review and analysis of co-creative systems in Graphic Design, evaluating them according to the fundamentals of Design, such as its principles and how they are applied.

## 3 METHOD

We provide a detailed description of the methodology employed in data collection and the dimensions defined to analyse the surveyed literature. Details of the corpus are available here: <Link>

### 3.1 Data Collection

The approach used was based on the PRISMA methodology [60], which is structured into six phases (see Figure 1).

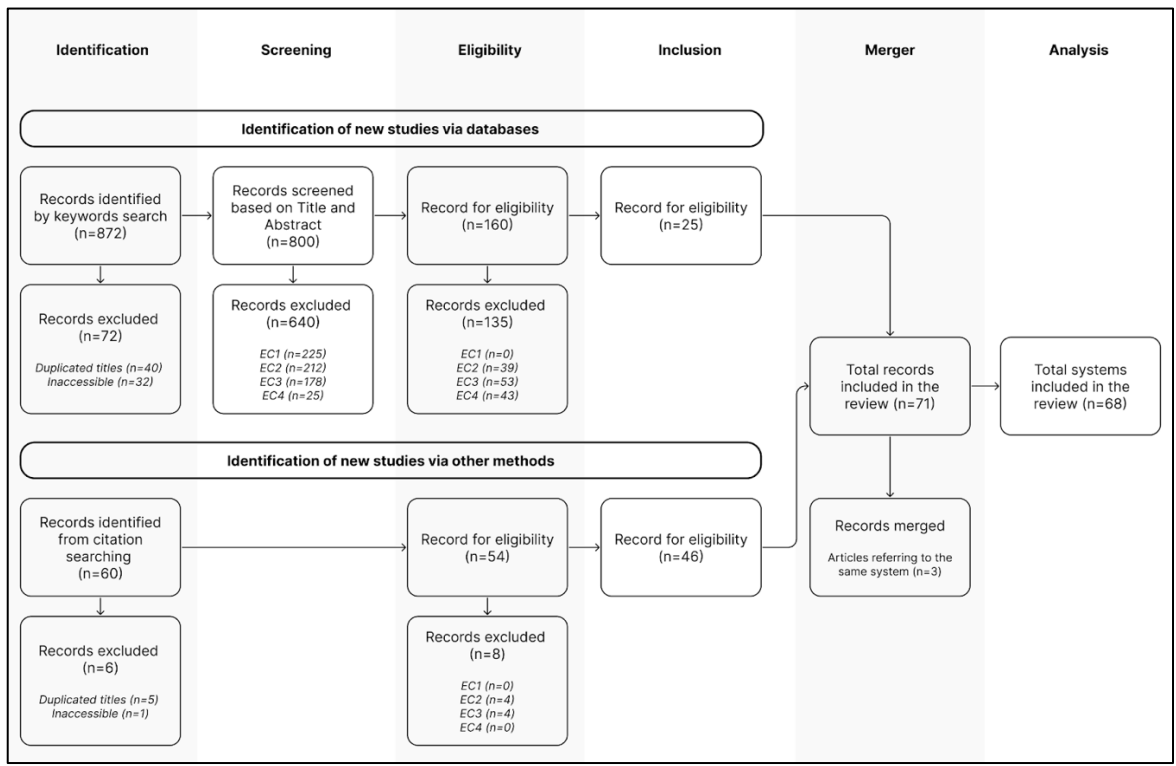

Figure 1: The process of data collection following the PRISMA framework.

*3.1.1 Identification*

This literature review aims to map the landscape of human-computer co-creative systems in Graphic Design. We focused on systems developed by researchers rather than commercially available tools. This decision was made considering that research contributions provide detailed tool descriptions, enabling more effective assessment [20].

The search query was defined following an initial study to identify relevant terms in the areas involved. To validate their suitability for the research focus, we identified some papers aligned with our objective and searched for these terms in the authors' titles, abstracts or keywords.

After testing various search queries to determine which combinations got the most results, we opted for the search query: 'graphic design' + creative system + co-creative system + creativity support tools. We utilised the Google Scholar database, as it offers a comprehensive overview of the current state of the art. The last search was conducted in October 2024. Our search produced 872 results. Articles with duplicate titles (40) and inaccessible articles (32) were removed, resulting in 800 articles for screening. Additional articles were identified through direct citations from the previously obtained items. We collected 60 articles, of which six were excluded due to duplication or inaccessibility.

*3.1.2 Screening, Eligibility and Inclusion*

Once the results were identified, we defined the following exclusion criteria to guide the selection and eligibility phases:

EC1: The article discussed design in a broad scope.
EC2: The article did not describe Graphic Design tasks.
EC3: The primary contribution of the paper was not related to creative systems.
EC4: The creative system for Graphic Design is not the main topic; instead, the article discusses designing creative systems.

The selection phase resulted in 160 articles for analysis. After the eligibility phase, 25 articles were included (see Table 1). In addition, 46 articles were included through the citation search, bringing the total to 71. During analysis, we found three articles referring to systems already represented in other articles; therefore, 68 unique systems were included in the final analysis. For all subsequent evaluation phases, the criteria remained the same.

Table 1: Classification of article references, distinguishing between the systematic review corpus and additional supporting citations.

| Type of Source | References |
| --- | --- |
| Systematic Review | [7, 10, 12, 15, 17, 21, 23, 24, 28, 29, 43, 47, 48, 52, 53, 62, 66, 67, 69, 72, 87, 88, 92, 94, 102] |
| Additional | [4–6, 8, 9, 11, 13, 16, 19, 22, 26, 27, 31–34, 37, 39–42, 46, 50, 56, 58, 61, 63, 73, 75, 79–86, 93, 95–101, 103] |

## 3.2 Data Analysis

Faced with the corpus collected, we defined a set of evaluation dimensions to guide our analysis. To ensure their relevance, we first reviewed a small, random sample of ten papers to identify structural tendencies. Considering the main objective and after several rounds of discussion and refinement, we defined three dimensions: (i) Collaborative Panorama—the nature of collaboration between human and system; (ii) Processes and Modalities—the processes and modalities explored during interaction; (iii) Graphic Design Principles—the integration of Graphic Design principles. These dimensions form the basis of our framework, which is presented in detail in the following section.

## 4 THE GRAPHIC FRAMEWORK

Under defined dimensions, we developed a framework (see Figure 2) that defines relevant parameters within the context of co-creative systems in Graphic Design, grounded in prior literature [30, 71, 78, 89]. Given the complexity of balancing scientific rigour with the subjective nuances inherent in the design field [44], this model serves as a conceptual guide for creators in developing effective and comprehensive co-creative systems.

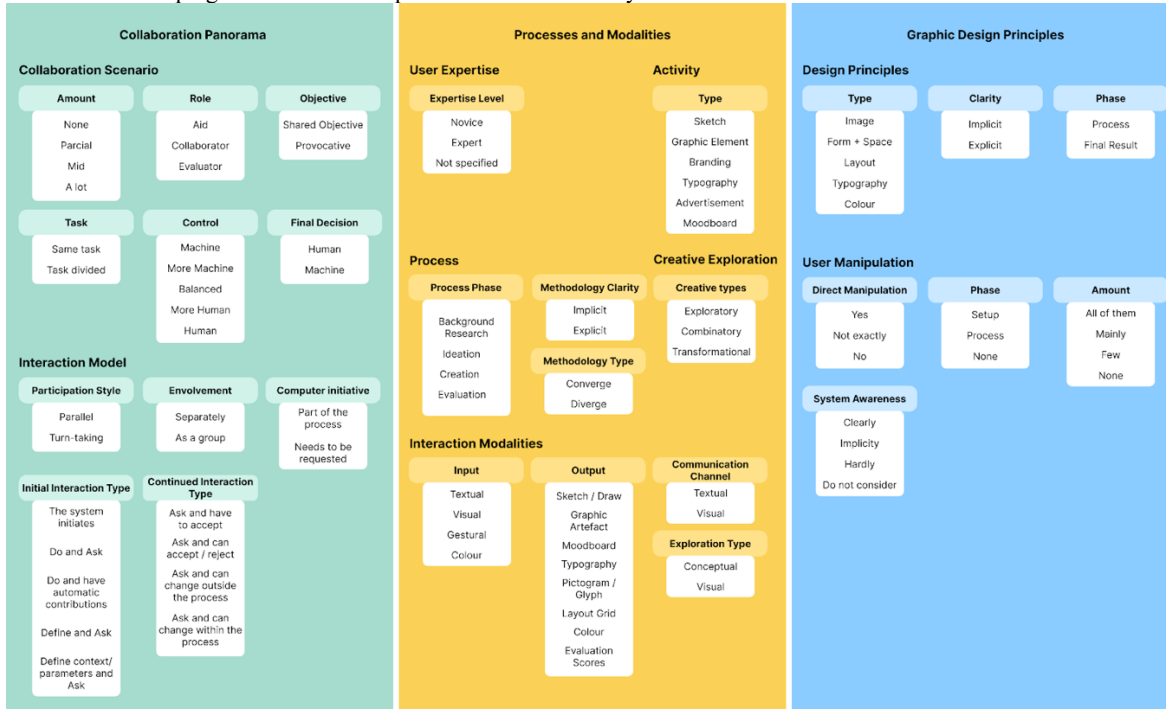

Figure 2: Overview of the GRAPHIC framework's three dimensions and key parameters.

*4.1.1 Collaboration Panorama*

This first dimension refers to human-computer collaborative panorama, focusing on the collaborative scenario and interaction model. The first category aims to understand the level of collaboration within the design process, the system's role, and who controls the process [78]. Concerning the level of collaboration and control, we defined options that reflect varying degrees of both aspects throughout the process. Regarding the system's role, we identified three distinct roles: Aid—the system acts as a helper; Collaborator—the system acts as a partner; and Evaluator—the system only evaluates the work done.

We also analysed the collaborative dynamics by examining the computer's objective when making its contributions, the division of tasks between the parties and who makes the final decision—human or machine [57]. Concerning the objective, we defined two possibilities: the computer follows the same objective as the human or acts provocatively [30, 36, 71, 76]. Concerning task division, we distinguished two scenarios: either agents work on the same task or work on different tasks [30, 36, 71].

The second category explores human-computer interactive dynamics. Based on the literature [30, 71], we analysed how contributions are organised, whether through turn-taking or in parallel, and the nature of involvement, whether both parties contribute to a shared environment or maintain separate environments. Additionally, we analysed the interactive dynamics in terms of system initiative and the nature of both initial and ongoing interaction. Regarding the first point, we analysed whether the system needs to be prompted (the user proceeds independently and requests input from the system when needed) or whether their input is automatically part of the process.

The type of initial interaction was analysed considering five categories: The system initiates—the process begins with a contribution from the system; Do and Ask—users contribute, notifying the computer that they have finished and asking for a response; Do and have automatic contributions—users make a first contribution and the computer automatically suggests recommendations; Define and Ask—users define all or most of the parameters and ask for a contribution; Define context/parameters and Ask—users give the computer clues or context and ask for a contribution.

As for the type of continuous interaction, we have defined four scenarios: Ask and have to accept—after requesting the computer's contribution, users must accept it without alternatives; Ask and can accept/reject—users can reject or accept the computer's contribution; Ask and can change outside the process—users are given access to an editable version of the output, allowing modifications, but outside of the collaboration environment; Ask and can change within the process—users can modify the computer's contribution during the process, which then continues based on those changes.

*4.1.2 Processes and Modalities*

This dimension aims to understand how design processes are structured and how their variants are addressed. The first two parameters, User Expertise and Activity Type, prove to be determining factors in shaping the process. People with different levels of expertise tend to have different needs [49], as do the types of activities.

To understand which phases these systems commonly support, based on existing literature [20, 78], we defined five phases that constitute the process, regardless of the task. In terms of methodology, we assessed the clarity of it, whether there is an explicit division between process phases, and whether it supports divergent and/or convergent thinking [78]. We also analysed the creative types explored, adopting Boden's [3] classification: combinatory, exploratory and transformational.

The third category addresses interactive modalities, including the system's inputs and outputs, communication channels, and exploration types. For inputs, we identified four modalities: text, visual (e.g., drawings), gesture and colour. Regarding communication channels and exploration types, we identified two categories for each. Communication channels can be visual or textual, while the exploration can be visual and conceptual (systems that operate on the semantic relationships between concepts).

*4.1.3 Graphic Design Principles*

One way to mitigate the challenges in Graphic Design is to utilise its principles as a communication platform. This dimension aims to understand the current systems in this context.

For the design principles parameter, we first analysed how different designers propose and categorise theoretical principles. The selection of principles was conducted systematically by identifying and comparing those advocated by various authors [45, 55, 68, 74, 89, 91]. After a comparative analysis, we chose to base our framework on White's [89] categorisation (Image, Form + Space, Layout, Typography and Colour), due to its structured organisation, which facilitates consistent evaluation across diverse systems.

In addition to identifying which principles are addressed, the framework also aims to understand how they are addressed during the process. This may occur either explicitly, the system presents the design principles clearly and identifiably as part of its operation or interface; or implicitly, where the system incorporates the principles during the creative process without naming or visibly presenting them to the user. Another aspect to consider is when these principles are addressed: during the creative process, by the system or the user, or only in the final result, only through the result can we see that the principles have been worked on during the process.

Given the central theme of this survey, it was also relevant to assess the degree of user control over the principles and the flexibility of that control. For the first parameter, we defined three levels: Yes—the user has direct control over the principles; Not exactly—the user can influence the principles, but not entirely directly; No—the user has no control over the principles. Regarding when this manipulation occurs, we identified two possible phases: during configuration or during the creative process.

## 5 RESULTS

In this section, we present the results obtained from the evaluation, both in terms of general information and human-AI collaboration, considering the three defined dimensions. Regarding general information, Figure 2 shows the distribution of articles over the years. A significant peak occurs in 2020, with a progressive increase since 2015. Despite a slight decline after 2021, the number of articles remains relatively constant until 2024. This trend reflects ongoing interest in creative systems applied to Graphic Design.

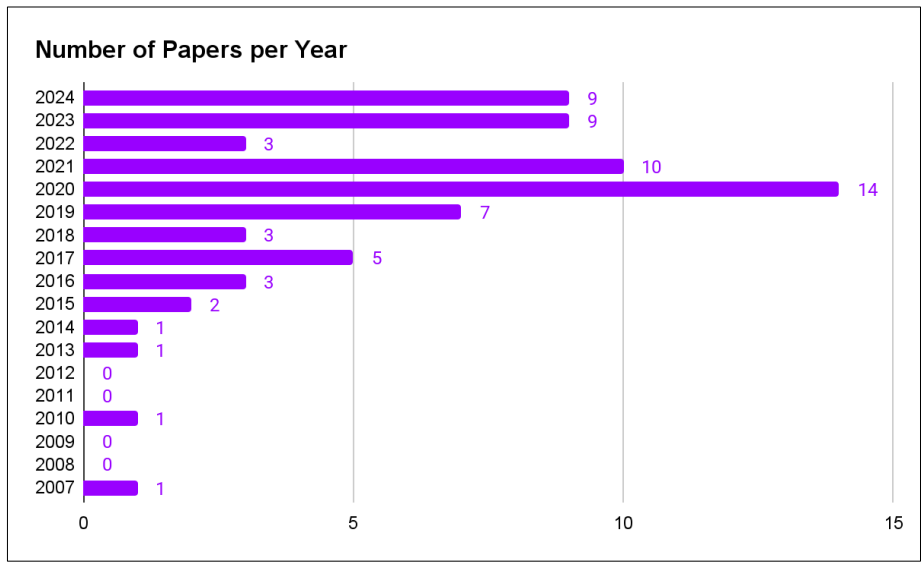

Figure 3: Frequency of papers per year.

### 5.1 Understanding Collaboration Panorama

To understand collaboration, we identified relevant evaluation parameters in human-computer interaction, including the collaboration scenario and interaction modalities.

*5.1.1 Collaboration scenario*

The corpus analysis reveals that human-computer collaboration scenarios are often limited and unbalanced. As shown in Figure 4, many of the analysed articles fall into the "Partial" collaboration category [5, 39, 40, 103], indicating that many systems still assign computational systems a predominantly supportive role. This is the case with the Evotype [58], a generative tool that creates glyphs exploring the boundaries between legibility and expressiveness. The system receives as input an SVG file containing a set of shapes used to develop the glyphs. The tool acts as an aid system with a partial level of collaboration.

Although a substantial number of articles are categorised as "Mid" [7, 34, 42], or "Significant" [22, 43, 88] collaboration, these remain fewer than those in the 'Partial' category. This suggests that many state-of-the-art systems do not yet promote a truly balanced human-computer partnership. This imbalance is further reflected when considering control during the creative process. Only 18 (26.47%) cases demonstrate equal levels of control between the user and the system, while in most cases (50, 73.53%), the user retains final decision-making authority [12, 16, 62].

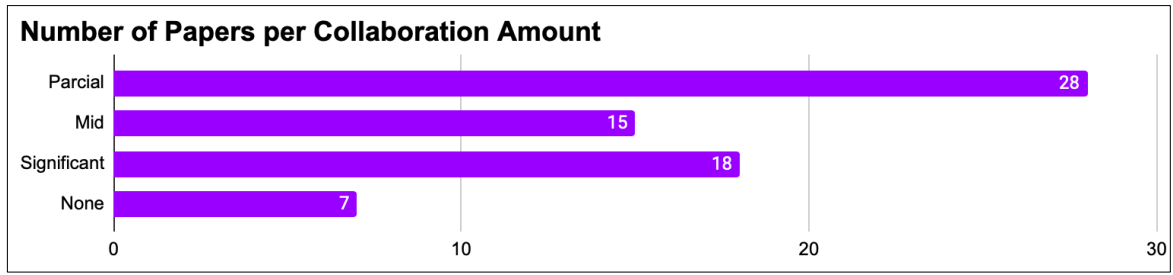

Figure 4: Frequency of papers per collaboration amount.

Another key factor is explainability. Although a few projects (8, 11.76%) try to implement strategies to mitigate explainability barriers [7, 29, 41], the vast majority (60, 88.23%) do not [9, 63, 96]. The Creative PenPal [31], a co-creative system that presents sketches to inspire users during design tasks, attempts to explain which object informed its suggestions to improve collaboration. Similarly, MagicBrush [92], a symbol-based Chinese painting system designed for beginners, provides explanations of the meanings of the selected elements. These systems only provide brief explanations of the contribution, not the process that led to it.

Most systems feature divided tasks with turn-taking [13, 15, 92] (see Figure 5), meaning that the human and the computer contribute at different times. Collaborative work on the same task is poorly represented [16, 28, 61]. For instance, VLT [87], a system for transferring layouts using vector graphics, enables both parties to share the same task and level of control, allowing mutual modification of contributions.

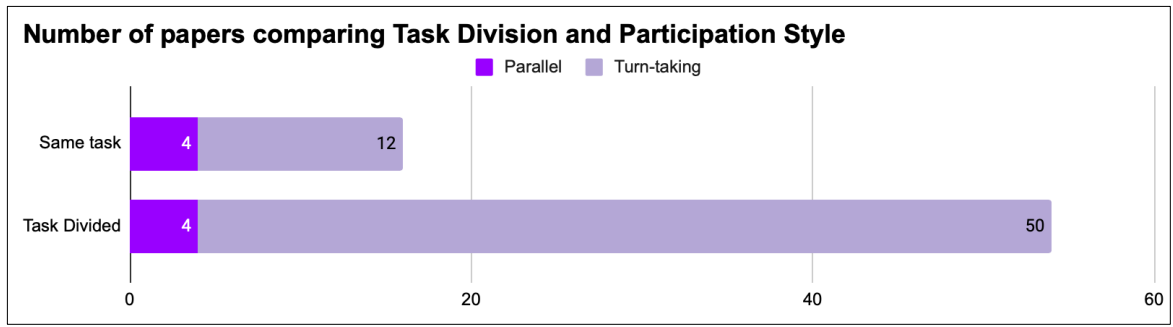

Figure 5: Frequency of papers by task division, confronted with participation style.

*5.1.2 Interaction model*

As mentioned before, the dominant participation style is turn-taking (61, 89.71%), with fewer examples of parallel interaction (8, 11.76%), and one case allowing for both styles of participation [62]. The robotic arm developed by Lim et al. [47] is an example of a system that allows parallel processes. The robotic arm enables both parties to draw independently yet toward a shared goal.

The analysis of initial interaction types (see Figure 6) shows that users most often lead initial interactions with the "Define parameters/context and ask" type [34, 75, 86], followed by "Define and Ask" [13, 96, 98]. In PatternPursuit [72], the user defines a theme, and the system selects images that will be worked on within that scope. In VisiFit [10], the user defines all the components, specifically the images they want to use, and the system generates a solution.

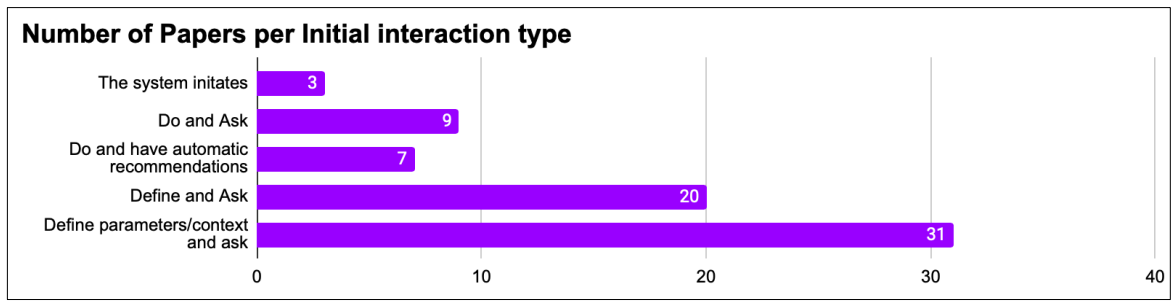

Figure 6: Frequency of papers per initial interaction type.

For ongoing interaction (see Figure 7), the most frequent type is "Ask and can accept/refuse" [15, 17, 21], in which users maintain control over the computer's contribution but do not necessarily influence the process. Cheng et al. [9] proposed a system that enables users to edit images interactively through a sequence of interactions, allowing them to explain their goals to the system. Cases where the user can negotiate or modify the system's contributions within the process are the second least frequent scenario [7, 22, 92], indicating that there is still room for the development of more collaborative and adaptive systems. ICONATE [101] addressed this by allowing users to directly modify the system's contribution, thereby generating more solutions in response to that change.

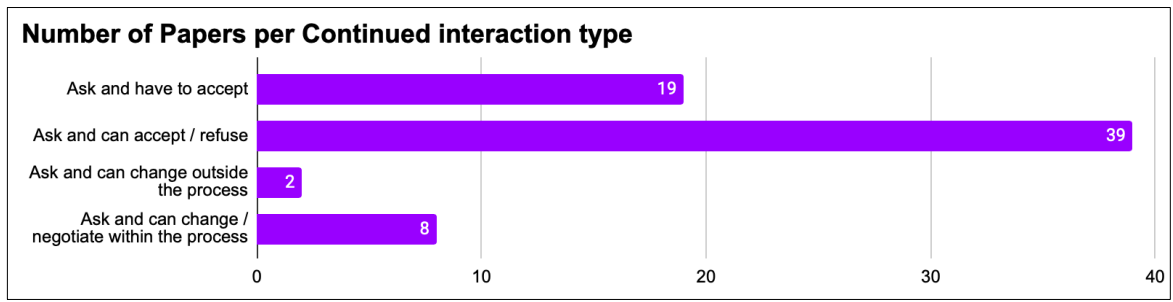

Figure 7: Frequency of papers per continued interaction type.

## 5.2 Understanding Exploration

This analysis reveals how creative systems are used to support activities within the graphic design discipline, particularly in terms of input/output modalities and coverage in supporting the design process.

*5.2.1 Target audience and activities*

Most systems were developed for Graphic Design specialists [12, 33, 34] (42, 61.76%), with 16 (23.53%) articles not specifying the target audience [32, 37, 43] and 10 (14.71%) developed for novices [6, 48, 92]. This suggests that many existing tools require prior knowledge of design practice, possibly because they include features that presuppose technical expertise. Warner [87] states that designers often draw on existing compositions, adapting them to their design. In this sense, the system he developed enables layout transfer by extracting the rules defined in the input layout and applying them to a given output design. This method also allows iterative refinement and negotiation of the rules defined in a multimodal way, from automatic rule standardisation to manual manipulation of elements. This approach allows the designer to directly apply various rules within the layout design principle, but requires some knowledge of their application.

As Figure 8 illustrates, most systems focus on exploring Graphic Elements/Artefacts [24, 27, 37]. This may indicate that the research landscape focuses more on generating individual graphic components that can be used in various contexts rather than on the application of these elements in specific contexts, such as advertising [22, 84] or branding [28, 102].

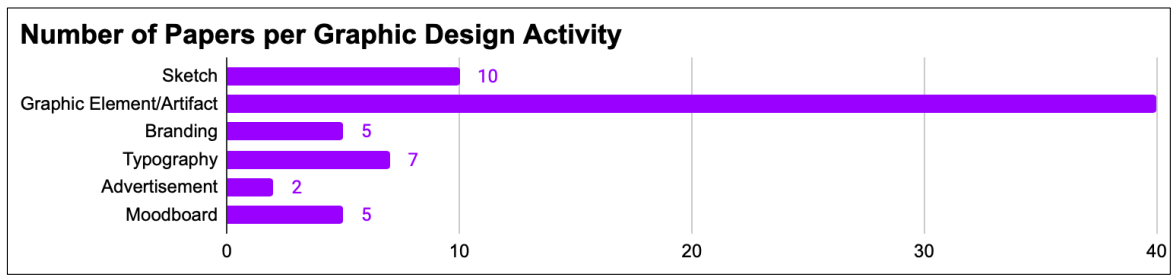

Figure 8: Frequency of papers per Graphic Design activity.

*5.2.2 Interaction Modalities*

In terms of interaction modalities explored in the corpus (see Figure 9), the visual input alone (27, 39.71%) [13, 19, 48] and the combination of visual + textual (22, 32.35%) [46, 61, 80] are the most common. Textual input (14, 20.59%) [11, 15, 41] also appears frequently. Likewise, visual channels are the most used for communication (52, 76.47%). This reflects the inherently visual nature of the Graphic Design discipline.

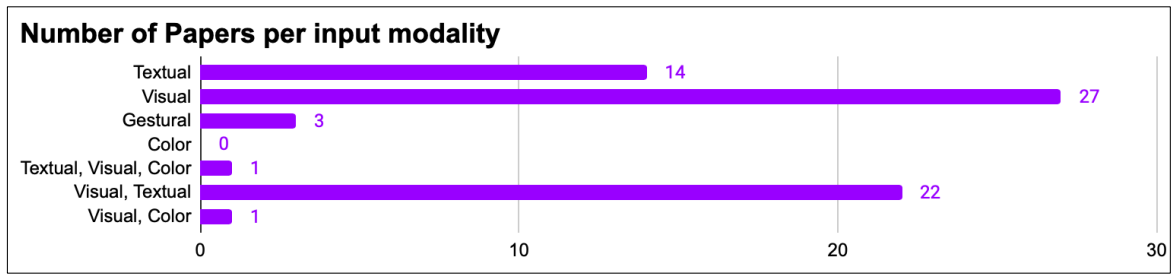

Figure 9: Frequency of papers per input modality.

All systems analysed employed visual exploration (68, 100%), and a smaller subset also supported conceptual exploration (29, 42.65%) [29, 43, 79]. This suggests interest in integrating semantic associations into AI-supported creativity, mirroring humans' creative reasoning.

Another relevant aspect concerns the type of results created by these systems. The data show a clear emphasis on traditional graphic artefacts, such as posters (16, 23.53%) [52, 88, 102], followed by images (12, 23.53%) [9, 24, 93] and sketches (12, 23.53%) [31, 47, 48]. Other outputs, such as typefaces (7, 10.29%) [85, 97, 103], mood boards (5, 7.35%) [12, 23, 41], and pictograms (4, 5.88%) [15, 37, 80], among others, also appear, but less frequently. There is also a small group of systems that focus on generating evaluation metrics (4, 5.88%) [82, 95, 98] or visual heat maps (3, 4.41%) [4, 19, 100], which tend to be focused on evaluating Graphic Design artefacts.

*5.2.3 Design process and creative exploration*

To understand the general panorama of the systems regarding the design process, we analysed the methodology applied to the collaborative processes conducted. Given this, we observed that only nine (13.24%) [12, 34, 72] systems have an explicit design process, while 59 (86.76%) [16, 26, 83] work with an implicit methodology. This indicates that most systems do not clearly separate the convergent and divergent phases of the process, making it difficult for the user to realise which phase they are in. Nonetheless, 39 (57.35%) systems support both convergent and divergent thinking [22, 62, 101], 16 (23.53%) support only convergent thinking [5, 63, 94], and three (4.41%) support only divergent thinking [16, 29, 41].

As Figure 10 shows, the corpus analysed focuses more on the development of systems that support the Creation phase [9, 28, 87], followed by the Ideation phase [8, 16, 61]. These stages often explore all three types of creativity defined by Boden [3], whereas the Background Research [29, 42, 72] and Evaluation [19, 82, 95] phases mainly support combinatorial and exploratory creativity. We can conclude that Transformational creativity is little explored, indicating that systems do not promote changes in the conceptual space during the creative process, acting as provocative partners.

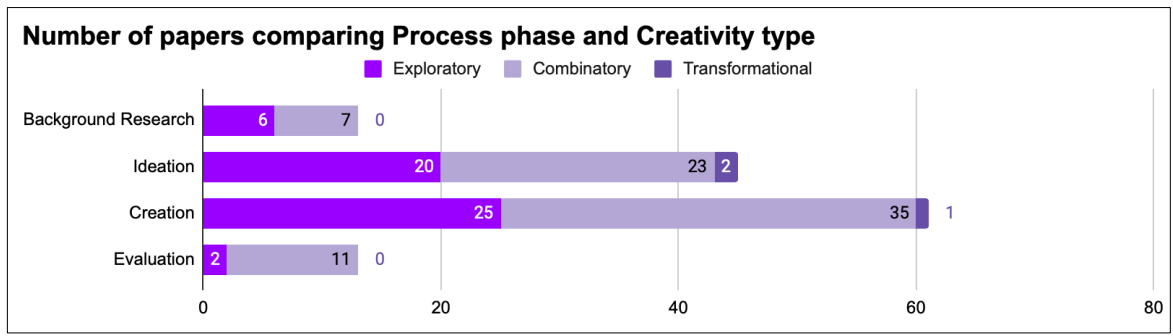

Figure 10: Frequency of papers confronting the design process phase and the creative type explored.

## 5.3 Understanding Graphic Design Principles

This dimension analyses how design principles are applied in systems and the user's capacity to engage with them. One of the aspects assessed was which of the selected principles are most used and how these systems allow the user to explore them directly.

*5.3.1 Design Principles*

The Graphic Design systems should function based on their fundamental principles to facilitate a graphic communication channel between humans and computers. As Figure 11 shows, the most worked on principle is Layout (37, 54.41%) [22, 52, 62], followed by Colour (34, 50%) [21, 40, 94] and Form+Space (32, 47.06%) [31, 37, 75]. These principles are mostly worked on implicitly (51, 75%) [8, 58, 100] and less explicitly (17, 25%) [22, 28, 87], and are worked on during the creative process (43, 63.24%) [16, 81, 92], rather than just in the final result (26, 38.24%) [37, 39, 100]. CreativeConnect [12] is an example of a system that operates on principles implicitly during the creative process. This system supports graphic designers in the ideation process by extracting elements from reference images. This way, the user can prioritise what the system should consider in the selected images, whether Layout or Form. With this information, the system generates sketches that follow the references placed on the mood board by the user and the respective requirements.

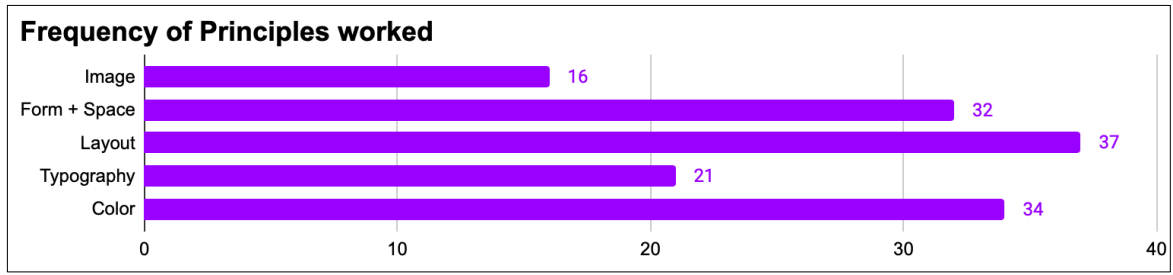

Figure 11: Frequency of papers per Graphic Principle worked on.

*5.3.2 User Manipulation*

Figure 12 reveals that in systems which apply design principles implicitly, users typically lack direct control and can only indirectly influence the rules applied [5, 19, 75]. Figure 13 shows when such influence occurs. In the case of direct manipulation, this occurs above all during the creative process [21, 43, 87]. When users can only influence indirectly, they can do so both during setup [39, 56, 83] and during the process [7, 29, 62]. There is a system that allows indirect manipulation during configuration and explicit manipulation during the process [22]. Vinci [22] is a collaborative system for generating advertising posters. In this sense, users can provide images of a product and its slogans and only choose the working elements, but cannot manipulate them. The system generates results based on existing posters, and users can then edit the poster and update the generated results to reflect their design preferences, thereby manipulating the design principles explicitly.

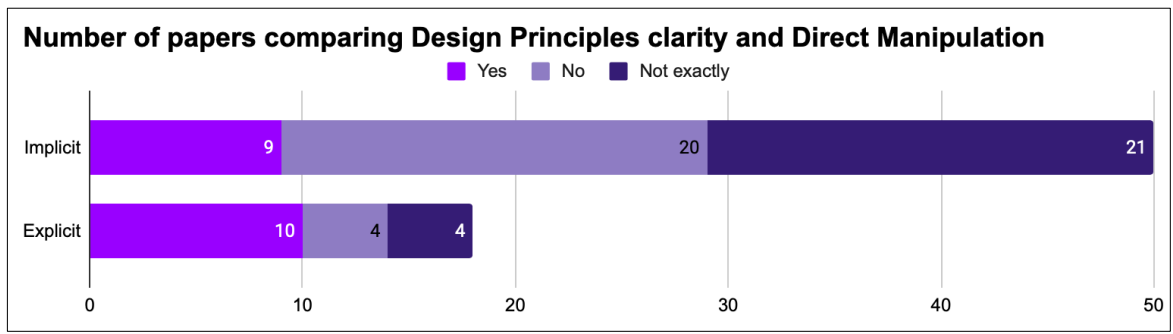

Figure 12: Frequency of papers by clarity of Design Principles, confronted with direct user manipulation.

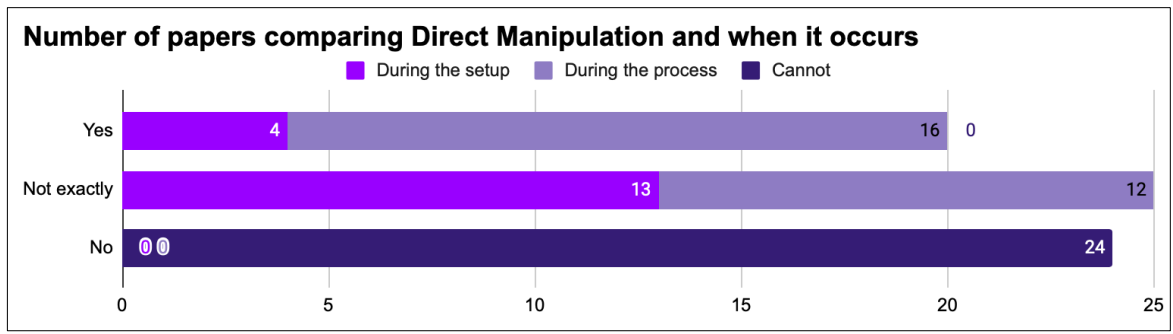

Figure 13: Frequency of papers by direct manipulation, confronted with the phase it occurs.

More systems do not consider (28, 41.18%) [11, 47, 94] or hardly consider (11, 16.18%) [7, 37, 92] how users want to work with each principle than those that do so implicitly (24, 35.29%) [15, 34, 75] or clearly (5, 7.35%) [22, 87, 88]. Despite being based on design principles, they tend not to learn from the user's choices. When they do, it is implicitly, with no clear cause-and-effect relationship or negotiation on how principles are applied. MetaMap [34] is an example of a system that implicitly aligns its contributions with user preferences regarding the principles. The convergence is not transparent, and the user lacks direct control, making it difficult to express their desired manipulation of the principles clearly.

By cross-referencing these values with the collaborative amount (see Figure 14), we see that systems with higher collaboration levels enable more balanced and diverse application of design principles [22, 28, 102]. This indicates that more robust systems in terms of collaborative scenarios not only increase the complexity of the interaction model but also manage to work more extensively with the different design principles. For instance, OptiMuse [102], a system for developing slides, allows users to input commands and engage in rule-based conversations to generate alternatives until their goals are met. This system works on all five principles selected.

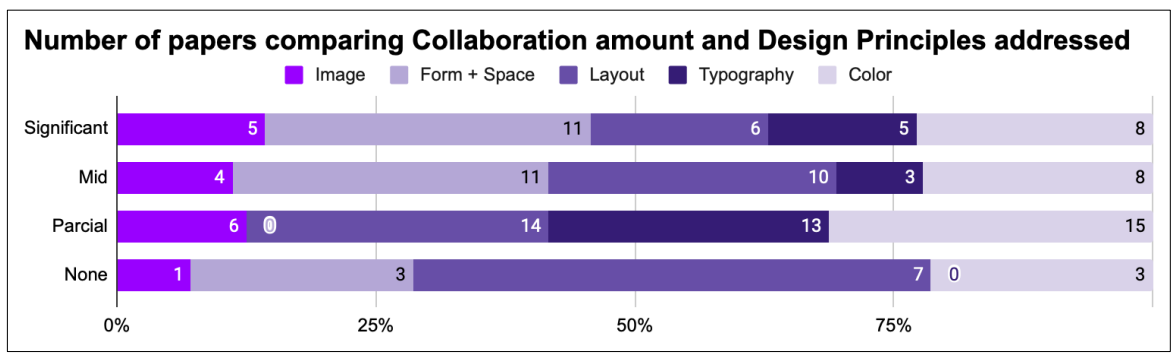

Figure 14: Design principles addressed according to the amount of collaboration.

Despite addressing design principles, users are often unable to manipulate them directly. Systems with high collaboration levels more frequently allow users to manipulate the design principles consciously (see Figure 15). Reframer [44], is a human-IA design system that supports an iterative and real-time process, and users can manipulate all the principles worked out by the system. At the "Partial" and "Mid" levels, most interactions fall under "No" or "Not exactly" categories, revealing limited users' autonomy. At the 'None' level, manipulation does not occur, indicating an almost total absence of practical influence on the fundamentals of Graphic Design. Figure 16 complements this conclusion by analysing when the user can influence the principles. It shows that greater human-computer collaboration tends to shift the user's influence from the setup phase to the creative process itself.

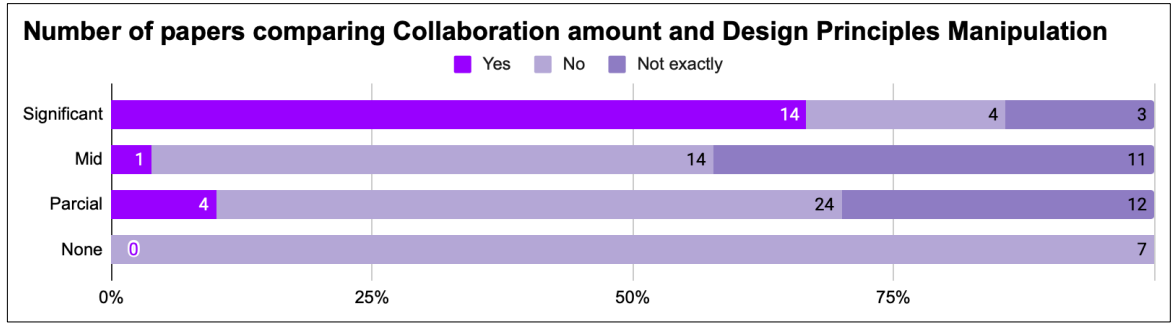

Figure 15: User's direct manipulation of design principles according to the amount of collaboration.

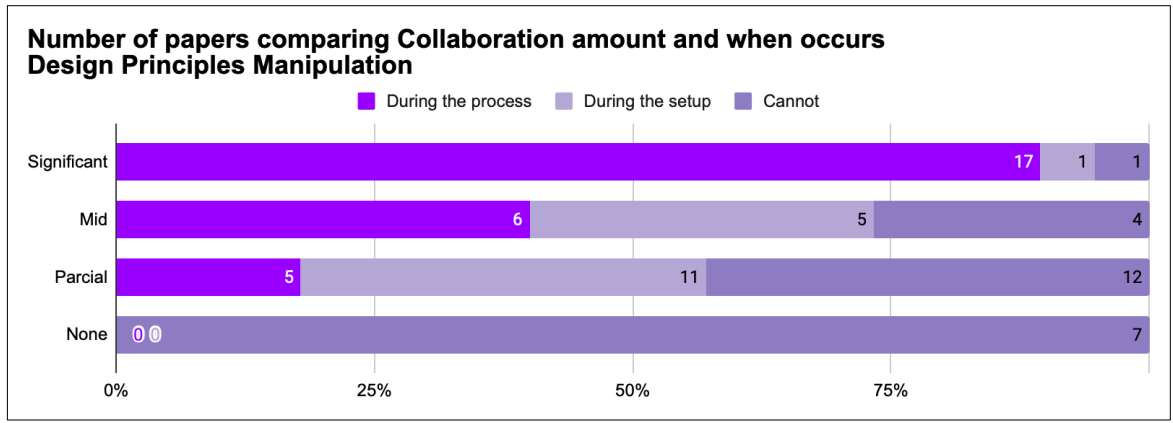

Figure 16: User's direct manipulation of design principles phases according to the amount of collaboration.

This data suggests a correlation between collaboration level and balanced manipulation throughout the design process. This reinforces the importance of designing systems that enable active, ongoing, and meaningful collaboration to enhance both computational and user creative potential in the design context.

## 6 DISCUSSION

The works presented in Section 5 represent a fair sample of co-creative system approaches applied to Graphic Design. Based on the analysis, several research opportunities were identified:

- Unbalanced collaboration: Systems often lack a balanced collaboration process, where human and computer share equal control. There is a lack of flexibility in task division, as tasks are predefined and cannot be altered. This lack of negotiation can influence the success of the process, as different users may have different needs in different aspects of the design process.

- Unbalanced initiative: Initiatives tend to be unbalanced, with the computer often acting when prompted. In contrast to human-human collaboration, we realise that what would be expected is for the two to intervene either equally or at least without needing to be called upon [18].
- Limited explainability: Few systems employ techniques that promote explainability. Although some systems include strategies, these remain insufficient, given the general panorama of the area. Design is highly subjective, making explainability fundamental in collaboration, as understanding our colleagues' reasoning helps us agree with it or improve decisions. In design, the concept gives strength to design decisions and, without an explanation, these can go unnoticed.
- Partial support for the design process: Most systems focus on specific phases, mainly ideation and creation, neglecting the design process as a whole, which extends from initial research to perfecting the proposed solution. There are a few systems that promote this collaboration throughout the process, reflecting a fragmented view of the creative process. This view limits the potential for collaboration and the development of coherent solutions.
- Narrow creativity types: Systems tend to support combinatorial and exploratory creativity, rarely promoting transformational creativity. Most operate within defined conceptual spaces with limited capacity for reframing them. This contradicts the potential for collaboration in a scenario where the user could take advantage of conceptual changes and not work with a partner who tends to limit themselves to exploring a previously defined space.
- Unbalanced control over design principles: In many systems, users cannot directly manipulate Graphic Design principles, which are instead controlled by the system. Furthermore, when the user can manipulate, few systems clearly consider their options when developing the artefact. This unbalanced control jeopardises both the user's expressiveness and the positive use of collaboration.
- Lack of relation among principles: While some systems address one or two principles, considering the classification we have chosen, few integrate three or more in an interrelated manner. This limits their ability to produce cohesive visual compositions where principles influence each other. Systems that integrate and relate all principles could better mirror real-world design processes and foster effective human-computer collaboration.

We recommend that developers of co-creative Graphic Design tools consider these shortcomings. The GRAPHIC can become a valuable evaluative tool for assessing such systems and promoting the advancement of the state-of-the-art. This is not closed, and dimensions other than those discussed here can be added, for instance, the usability of systems. An interface that supports the interaction models mentioned in a user-friendly way is essential for successful human-computer collaboration.

## 6.1 Limitations

One limitation of this research is that the systematic review is based solely on the Google Scholar database. However, given the interdisciplinary nature of Graphic Design and co-creative systems, Google Scholar was selected for its broad indexing capabilities, which helped to capture relevant publications from diverse academic communities that might have been missed in more domain-specific databases.

## 7 CONCLUSION

This article presents a systematic review of 68 systems published between 2007 and 2024 on human-computer collaboration in the field of Graphic Design. This review fills a significant gap in understanding how computational systems should effectively support Graphic Design processes. The analysis identifies research opportunities, including the need to balance collaboration and initiative on both sides, improve communication between designers and AI through explainable approaches, and develop systems that focus more on transformational creativity and are based on established design principles. Another opportunity identified was to develop systems that support the design process continuously, from the exploration phase to the finalisation of the artefact.

This literature review provides a framework to promote the development of more effective computational systems in the design field, fostering creative, flexible, and realistic collaboration between humans and computers. The framework can serve as a valuable evaluation tool for advancing the state of the art and can be expanded with complementary dimensions, such as usability, which also plays a crucial role in the success of human-computer interaction.

## ACKNOWLEDGMENTS

This work is funded by national funds through FCT – Foundation for Science and Technology, I.P., within the scope of the research unit UID/00326—Centre for Informatics and Systems of the University of Coimbra. It is also funded by European Regional Development Fund (FEDER), through the Central Regional Program (Centro2030), Portugal 2030 and the European Union | Operation No 17372, Funds Branch Operation Code CENTRO2030-FEDER-01186800.

[103] Lian Zhouhui, Bo Zhao, Xudong Chen, and Jianguo Xiao. 2018. EasyFont: A Style Learning-Based System to Easily Build Your Large-Scale Handwriting Fonts. *ACM Trans. Graph.* Vol 38, No 1 (2018). Retrieved February 26, 2025 from https://dl.acm.org/doi/abs/10.1145/3213767**Prior Publication Statement**

This manuscript is original work and has not been published previously. It is not currently under consideration for publication elsewhere. This submission has no significant overlap with any other paper by the authors, either published or concurrently submitted.

26